\documentclass[fleqn,usenatbib]{mnras}

\usepackage[T1]{fontenc}

\DeclareRobustCommand{\VAN}[3]{#2}
\let\VANthebibliography\thebibliography
\def\thebibliography{\DeclareRobustCommand{\VAN}[3]{##3}\VANthebibliography}

\usepackage{graphicx}	% Including figure files
\usepackage{amsmath}	% Advanced maths commands
\usepackage{amssymb}	% Extra maths symbols
\newcommand{\update}{}
\newcommand{\scatterLOGVR}{0.14 dex}
\newcommand{\scatterLOGMSTAR}{0.46 dex}

\title[The stellar mass-binding energy relationship]{A universal relationship between stellar masses and binding energies of galaxies}

\author[Y. Shi et al.]{
Yong Shi,$^{1,2}$\thanks{E-mail: yshipku@gmail.com}
Xiaoling Yu,$^{1,2}$
Shude Mao,$^{3,4}$
Qiusheng Gu,$^{1,2}$
Xiaoyang Xia,$^{5}$
Yanmei Chen,$^{1,2}$
\\
$^{1}$School of Astronomy and Space Science, Nanjing University, Nanjing 210093, China.\\
$^{2}$Key Laboratory of Modern Astronomy and Astrophysics (Nanjing University), Ministry of Education, Nanjing 210093, China.\\
$^{3}$Department of Astronomy, Tsinghua University, Beijing 100084, China.\\
$^{4}$National Astronomical Observatories, Chinese Academy of Sciences, 20A Datun Road, Chaoyang District, Beijing 100012, China.\\
$^{5}$Tianjin Astrophysics Center, Tianjin Normal University, Tianjin 300387, People's Republic of China.\\
}

\date{Accepted XXX. Received YYY; in original form ZZZ}

\pubyear{2021}

\begin{document}
\label{firstpage}
\pagerange{\pageref{firstpage}--\pageref{lastpage}}
\maketitle

\begin{abstract}

 In this study we demonstrate that stellar masses of galaxies ($M_{\rm
   star}$) are universally correlated through a double power law function
 with the product of the  dynamical velocities ($V_{\rm e}$) and sizes
 to one-fourth  power ($R_{\rm e}^{0.25}$) of  galaxies, both measured
 at  the effective  radii.   The product  $V_{\rm e}R_{\rm  e}^{0.25}$
 represents  the fourth  root  of the  total  binding energies  within
 effective  radii  of  galaxies.   This  stellar  mass-binding  energy
 correlation has an observed scatter of \scatterLOGVR\, in log($V_{\rm
   e}R_{\rm   e}^{0.25}$)   and  \scatterLOGMSTAR\,   in   log($M_{\rm
   star}$). It holds for a variety of galaxy types over a stellar mass
 range of nine orders of  magnitude, with little evolution over cosmic
 time.   A toy model of  self-regulation between binding energies and
 supernovae feedback  is shown  to be able  to reproduce  the observed
 slopes, but the underlying physical mechanisms are still unclear. The  correlation can  be a potential  distance estimator
 with an uncertainty of 0.2 dex independent of the galaxy type.
  
\end{abstract}

\begin{keywords}
galaxies: general -- galaxies: haloes -- galaxies: evolution
\end{keywords}

%%%%%%%%%%%%%%%%%%%%%%%%%%%%%%%%%%%%%%%%%%%%%%%%%%

%%%%%%%%%%%%%%%%% BODY OF PAPER %%%%%%%%%%%%%%%%%%

\section{Introduction}

Galaxies form  in haloes of  dark matter.  The interplay  between dark
matters  and baryons  leads  to coupling  between  halo properties  and
galaxy  properties,  which  offers  crucial  clues  to  formation  and
evolution of galaxies.  The spatial  distributions of baryons on large
scales roughly  follow those of dark  matters because of the  force of
gravity.  This  results in a  relationship between halo masses  and galaxy
stellar masses, but such a correlation shows large scatters at low masses
as seen both in  observations and simulations \citep[e.g.][]{Miller14,
  Oman16,  Beasley16, vanDokkum18,  vanDokkum19, Read17,  Shi21}.  The
similar spatial distributions of baryons and dark matters also suggest
similar  tidal  torques  over the  cosmic  history  \citep{Peebles69},
leading to an  expectation that the specific  angular momenta (angular
momenta  per  mass)  of  galaxies  are  similar  to  those  of  haloes
\citep{Fall80, Mo98}, but observations show  that this is correct only
to  some  extent  \citep{Kravtsov13,  Posti18}. While  the  above  two
relationships   connect  galaxies   with  the   whole  haloes,   other
relationships  focus  more on  correlation  between  galaxies and  inner
haloes, such  as the  Tully-Fisher relationship of  late-type galaxies
\citep{Tully77},  the  Faber-Jackson  law  of  elliptical/S0  galaxies
\citep{Faber76}   and   the   fundamental   plane   of   elliptical/S0
\citep{Dressler87, Mobasher99}. However, we still lack a universal relationship
that holds  for all types of galaxies to probe connections between galaxies and (inner) haloes.

In this study, we compiled a  set of galaxies that
cover  diverse types  to demonstrate  that galaxy  stellar masses  are
universally correlated with the total binding energies within
  effective   radii  of galaxies.    A   cosmological   model   with   h=0.73,
$\Omega_{0}$=0.27  and $\Omega_{\lambda}$=0.73  is adopted  throughout
the study.

\section{The Compiled Galaxy Samples}

\subsection{General procedures to homogenize the dataset}\label{sec_samp_gen_proc}

The sample of galaxies includes almost all types of galaxies as listed
in Table~\ref{tab_stats}, with a total of 752 objects. 
We compiled measurements of  three physical
parameters, including  the galaxy  stellar mass ($M_{\rm  star}$), the
effective radius ($R_{\rm  e}$) of the galaxy, and  the dynamical
velocity at the  effective radius ($V_{\rm e}$) as  defined below.  In
order to have  accurate measurements of $V_{\rm e}$,  for the majority
of  galaxy types  in  this study,  we used  the  sample with  spatially
resolved maps of gas/stellar kinematics as offered by optical integral
field unit and radio interferometric observations. Because of this reason, early-type
galaxies of the  Sloan Digital Sky Survey (SDSS) are only used
as a sanity check (see \S~\ref{sec_fp}).

The  homogenization  of $M_{\rm  star}$:  all  measurements are  first
re-scaled  to our  adopted cosmological  model for  distances that  are
based on the Hubble flow. The  mass-to-light ratio is based on a fixed
set  of stellar  synthetic  spectra  \citep{Schombert19}.  We  adopted
$M_{\rm  star}/L_{3.6{\mu}m,  \odot}$=0.6  for both  star-forming  and
quiescent  galaxies, and  $M_{\rm star}$/$L_{B,  \odot}$=2.57, $M_{\rm
  star}$/$L_{V,   \odot}$=2.36,   $M_{\rm  star}/L_{R,   \odot}$=2.14,
$M_{\rm star}/L_{r,  \odot}$=2.14 and $M_{\rm  star}/L_{g, \odot}$=2.4
for quiescent galaxies by assuming  $B-V$=0.74.  An additional 0.1 dex
error is added quadratically to account for the systematic uncertainty
of the stellar mass \citep[see references in ][]{Shi21}.

The  homogenization of  $R_{\rm e}$:  similarly, all  measurements are
first re-scaled  to our adopted  cosmological model for  distances that
are based on  the Hubble flow. For quiescent galaxies, the
effective radii are corrected for  the projection effect following the
method in \citet{Wolf10}.

The homogenization  of $V_{\rm e}$: $V_{\rm  e}$ is the velocity  of a
circular orbit at  $R_{\rm e}$  due to the  total gravity from
dark matter, stars and gas. In star-forming galaxies where rotation dominates, $V_{\rm e}$
is   obtained   by   interpolating   the  rotation   curve.  In
quiescent galaxies where dispersion dominates,  $V_{\rm e}$  is based on  the dynamical
mass within the  effective radius through Newton's  law of gravitation
as $V_{\rm  e}$=$\sqrt{GM_{(R<R_{\rm e})}/R_{\rm e}}$.  If  the dynamical
mass is  not available, the light-weighted  velocity dispersion within
the effective radius is transformed to $V_{\rm e}$ by multiplying with
a  factor   of  $\sqrt{5/2}$ 
\citep[see   Equation  19  in  the   study  of][]{Cappellari06}.  If the velocity  dispersion is measured within
an aperture other than the  effective radius, the conversion is adopted
from the study  of \citet{Cappellari06} (see their Equation  1). For a
few measurements  where the aperture  is unavailable, we  just adopted
the published velocity dispersion.

\subsection{A summary of individual galaxy types}

Here a summary of  individual galaxy  types is presented. Some
details are also given about  how to homogenize the data for  a few studies that
we cannot adopt the above general procedure. Star-forming galaxies include
ultra diffuse  galaxies in  the field, low surface brightness galaxies,
spirals and irregulars, starburst galaxies, as well as high-z star-forming.
Quiescent galaxies include local  spheroidal galaxies, ultra compact  dwarfs,
ultra  diffuse galaxies in  clusters, dwarf elliptical, elliptical/S0, brightest cluster galaxies,
high-z quiescent galaxies and high-z massive compact galaxies.

1. Local  spheroidal galaxies:  for these  least massive  galaxies, we
included  objects found  around the Milky Way \citep{Wolf10}  and  M31 \citep{Tollerud12}.   Their   dynamical  masses were   obtained  by
modeling  the   line-of-sight  velocity  dispersion   of  individual
stars. Both  the effective radii and  stellar masses were based  on the
$V$-band photometry.

2.   Ultra compact  dwarfs: for  these smallest  galaxies, we  included
objects from \citet{Mieske08} and \citet{Forbes14}.  For both samples,
the  effective radii  and stellar  masses were  based on  the $V$-band
photometry.  The  kinematic measurements of two  samples were obtained
through  multi-object  fibers  and slitlets,  respectively.   For  the
sample  of  \citet{Mieske08},  we   assumed  half  of  their  measured
dynamical mass to be within the effective radius and converted them to
$V_{\rm e}$.  For  the sample of \citet{Forbes14},  we derived $V_{\rm
  e}$ from the central velocity dispersion following the above general
procedure.

3.  Ultra  diffuse galaxies in  the field:  we derived $V_{\rm  e}$ of
these  galaxies   from  the  rotation  curves   in  \citet{Shi21}  and
\citet{ManceraPina19}. We  further measured  the effective  radii from
the  $g$-band  images  of  the   Dark  Energy  Camera  Legacy  Surveys
\citep{Dey19}  following the  method  in  \citet{Shi21}.  Among  these
galaxies, AGC  242019 is also detected  in 3.6 $\mu$m that  is used to
derive its stellar mass. We then  estimated the mass to light ratio in
the $g$-band  of this  object  and  applied it  to  all remaining  diffuse
galaxies.

4. Ultra diffuse galaxies in clusters: the effective radii and stellar
masses  of  these  galaxies  were  based   on  the  $V$  or  $g$  data
\citep{Beasley16,   vanDokkum16,    vanDokkum18,   vanDokkum19}.    We
converted the dynamical masses within half-light radii to $V_{\rm e}$.

5.  Dwarf  ellipticals: for these  objects in  Virgo at a  distance of
16.7 Mpc \citep{Toloba12}, the stellar masses and effective radii were
measured  based on  the  $V$-band images.  We  converted the  velocity
dispersion within  the effective  radii to  $V_{\rm e}$  following the
above general procedure.

6. Low surface brightness galaxies:  the effective radii of objects in
\citet{deBlok01} were measured from the $B$ band images.  We converted
their $B$ band luminosities to stellar masses by adopting the  mass to light ratio
of  quiescent galaxies  because low  surface brightness  galaxies are
red.  We  estimated their  $V_{\rm e}$ from  the rotation  curves. Two
giant  low surface  brightness galaxies,  Malin  1 and  NGC 7589,  are
included  too.   We estimated  the  effective  radius  of Malin  1  by
combining both the bulge and  disk components in \citet{Bothun87}. The
effective radius  of NGC  7589 was  estimated by  \citet{Impey96}.  We
derived  stellar masses  of  both objects from the  $B$ band  luminosities
\citep{Lelli10} assuming old stellar  populations, and estimated their
$V_{\rm e}$ from their rotation curves \citep{Lelli10}.

7. Spirals  and irregulars:  for spirals  and irregulars  from Spitzer
Photometry and Accurate Rotation Curves (SPARC) \citep{Lelli16}, their
stellar masses  and effective radii  were measured in 3.6  $\mu$m.  We
estimated  $V_{\rm e}$  from  their rotation  curves.  For  additional
low-mass dwarf irregulars from  \citet{Read17}, we adopted the stellar
masses  and effective  radii  that were  estimated  from the  spectral
energy distribution  (SED) fitting \citep{Zhang12}.  We  derived their
$V_{\rm e}$ from their rotation curves.

8.   (Ultra) luminous infrared galaxies (LIRGs/ULIRGs):  for  these  objects \citep{Bellocchi13},  we
estimated their  $V_{\rm e}$ from  their dynamical masses  by assuming
that half of  masses are within the effective  radii.  Their effective
radii were measured in the near-infrared band. To estimate the stellar
mass, we  first extracted the  2MASS $J$  and $K$ photometry  from the
2MASS archive.   We then used  the result  of \citet{U12} to  obtain a
calibration  about the  mass-to-light ratio  in $K$  as a  function of
$J-K$  color.  Starbursts  galaxies  are highly  extincted; \citet{U12} accounted for this  by carrying out the full SED
fitting  from the  UV  to the  near-IR.  For  an  object with  several
counterparts,  the stellar  mass  was evenly  divided.   We adopted  a
factor of two larger systematic uncertainty for starburst galaxies.

9.   Ellipticals/S0:  for   galaxies  in   \citet{Cappellari13},  their
effective  radii were  based on  the combinations  of the  optical and
near-IR  images  \citep{Cappellari11}.   We  estimated  their  stellar
masses from the $r$-band luminosities,  and their $V_{\rm e}$ from the
velocity dispersion within the effective radii.

10.   Brightest    cluster   galaxies:    for   these    galaxies   in
\citet{Loubser08}, we  converted their central velocity  dispersion to
$V_{\rm  e}$.  We converted  their  $B$-band  luminosities to  stellar
masses. The effective radii are based on the 2MASS images.

11.  High-z quiescent  galaxies:  for objects  in \citet{Belli14},  we
converted the velocity dispersion within  the effective radii to $V_{\rm
  e}$. Their  effective radii  were measured in  F160W of the Hubble
Space Telescope (HST). Their stellar  masses were based on the fitting
to the optical-to-infrared SED. We homogenized them to the Kroupa IMF.

12. High-z  star-forming galaxies:  for galaxies  in \citet{Genzel17},
their   stellar    masses   were   obtained   by    fitting    the
optical-to-infrared SED, and we  homogenized them to the Kroupa IMF. The
effective radii were  measured in $H$ band. We  adopted their circular
velocities at the effective radii as $V_{\rm e}$.  For high-z low-mass
star-forming galaxies in \citet{Miller14}, we derived $V_{\rm e}$ from
the circular velocity at 2.2 times the disk scale length by adopting a
universal rotation  curve \citep{Persic96}. Their stellar  masses were
based on the fitting to  the optical-to-infrared SED, and we corrected
them to  the Kroupa IMF. The  effective radii were extracted  from HST
F125W images.

13.   High-z massive  compact  galaxies: two  such  galaxies are  from
\citet{vanDokkum09}  and   \citet{VanDeSande11},  respectively.  Their
stellar  masses were  based  on the  optical-to-infrared SED  fitting,
which  are corrected  to the  Kroupa IMF.  Their effective  radii were
measured from HST near-IR images.  We estimated their $V_{\rm e}$ from
the velocity dispersion within the effective radii.


\begin{table*}
\footnotesize
%\begin{center}
\caption{\label{tab_stats} Statistics of different galaxy types in Figure~\ref{V_R0dot25_Mstar}. }
\begin{tabular}{llllllllllllll}
\hline
Types of galaxies                 & \# of obj. & Median offset of log($V_{\rm  e}R_{\rm e}^{0.25}$)   & s.d. of log($V_{\rm  e}R_{\rm e}^{0.25}$) &  Reference   \\
                                  &            & (dex)            & (dex)   &  \\
\hline

Local Spheroids                     & 32  & 0.008   & 0.156 & 1,2 \\
Ultra Compact Dwarfs                & 31  & -0.018  & 0.130 & 3,4 \\
Ultra Diffuse Galaxies in field     & 5  & -0.002   & 0.108 & 5,6 \\
Ultra Diffuse Galaxies in clusters  & 13 & 0.075   & 0.297 & 7,8,9,10,11 \\
Dwarf Ellipticals                   & 29 & -0.068  & 0.155 & 12,13 \\
Low Surface Brightness              & 14 & -0.032  & 0.108 & 14,15 \\
Spirals and Irregulars              & 186 & -0.051  & 0.139 & 16,17 \\
LIRGs/ULIRGs                        & 39 & -0.085  & 0.182 & 18 \\
Elliptical/S0                       & 258 & 0.048   & 0.096 & 19 \\
Brightest Cluster Galaxies          & 40 & -0.009   & 0.090 & 20 \\
High-z Quiescent                    & 56 & 0.044   & 0.108 & 21 \\
High-z Star-Forming                 & 47  & 0.021   & 0.144 & 22,23 \\
High-z Massive Compact Galaxies     & 2  & 0.089   & 0.054 & 24,25 \\

\hline
\end{tabular}\\
References: 1-\citet{Tollerud12}, 2-\citet{Wolf10}, 3-\citet{Forbes14}, 4-\citet{Mieske08}, 5-\citet{ManceraPina19}, 6-\citet{Shi21}, 7-\citet{Beasley16}, 8-\citet{Chilingarian19}, 9-\citet{vanDokkum16}, 10-\citet{vanDokkum18}, 11-\citet{vanDokkum19}, 12-\citet{Forbes14}, 13-\citet{Toloba12}, 14-\citet{Lelli10}, 15-\citet{deBlok01}, 16-\citet{Lelli16}, 17-\citet{Read17}, 18-\citet{Bellocchi13}, 19-\citet{Cappellari13}, 20-\citet{Loubser08}, 21-\citet{Belli14}, 22-\citet{Genzel17}, 23-\citet{Miller14}, 24-\citet{VanDeSande11}, 25-\citet{vanDokkum09}
%\end{center}
\end{table*}

\begin{figure}
  \begin{center}
    \includegraphics[scale=0.35]{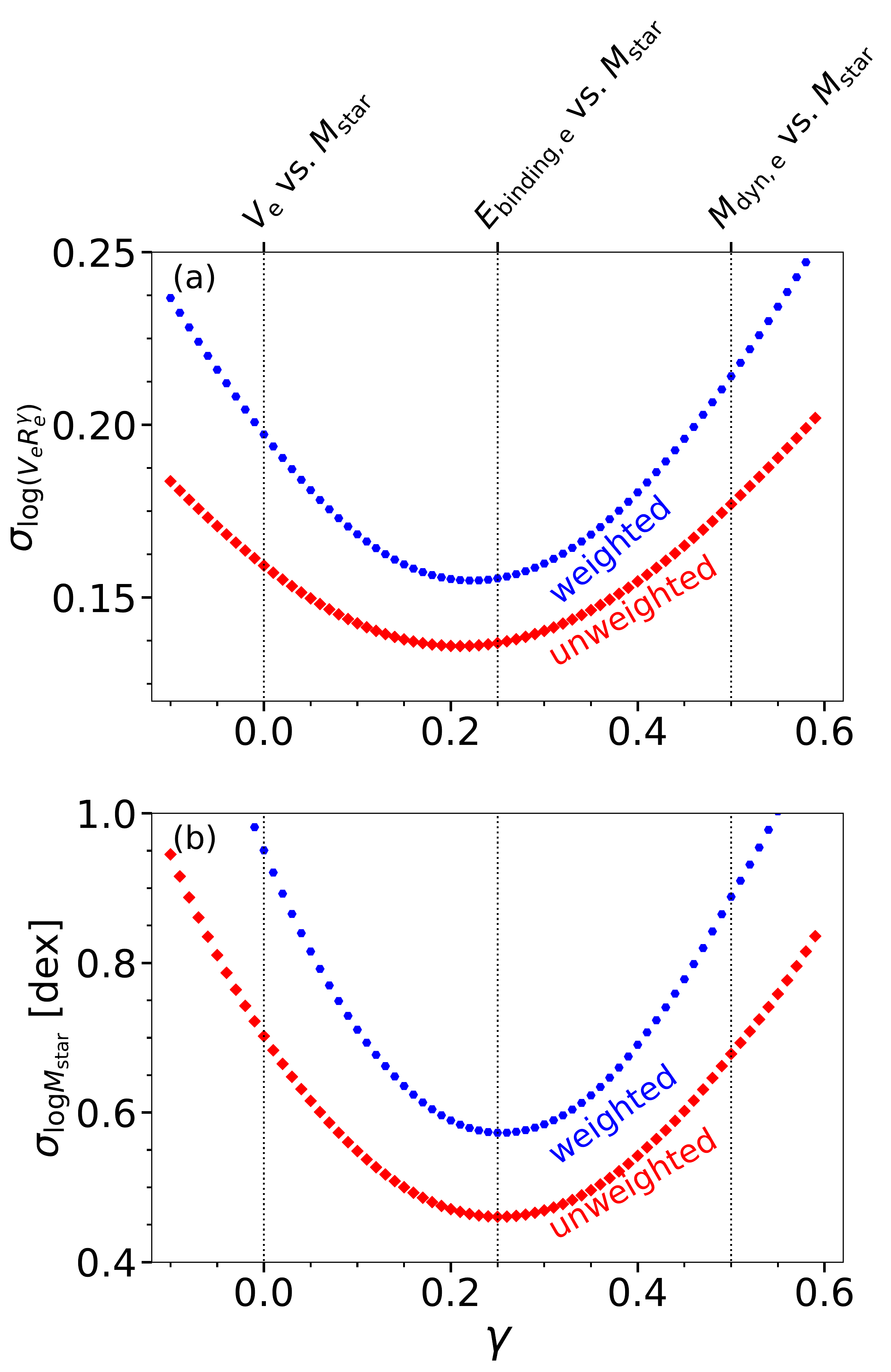}
    \caption{\label{dispersion_gamma} {\bf The dispersion of the best-fitted
        Equation~\ref{eqn_dpl_gamma} as a function of $\gamma$}. {\bf (a)} the dispersion
    in log($V_{\rm       e}R_{\rm        e}^{\gamma}$). {\bf (b)} the dispersion
    in log($M_{\rm star}$). The unweighted dispersion is the standard deviation, while
    the weighted dispersion is the one weighted by the number of each galaxy type (see text).
    Three dotted lines mark $\gamma$ values at which $V_{\rm       e}R_{\rm        e}^{\gamma}$
    represents the $V_{\rm       e}$, the binding energy within $R_{\rm e}$
    and the dynamical mass within $R_{\rm e}$, respectively, as illustrated
    in Equation~\ref{eqn_gamma_physics}. }
\end{center}
\end{figure}

\begin{figure*}
  \begin{center}
    \includegraphics[scale=0.6]{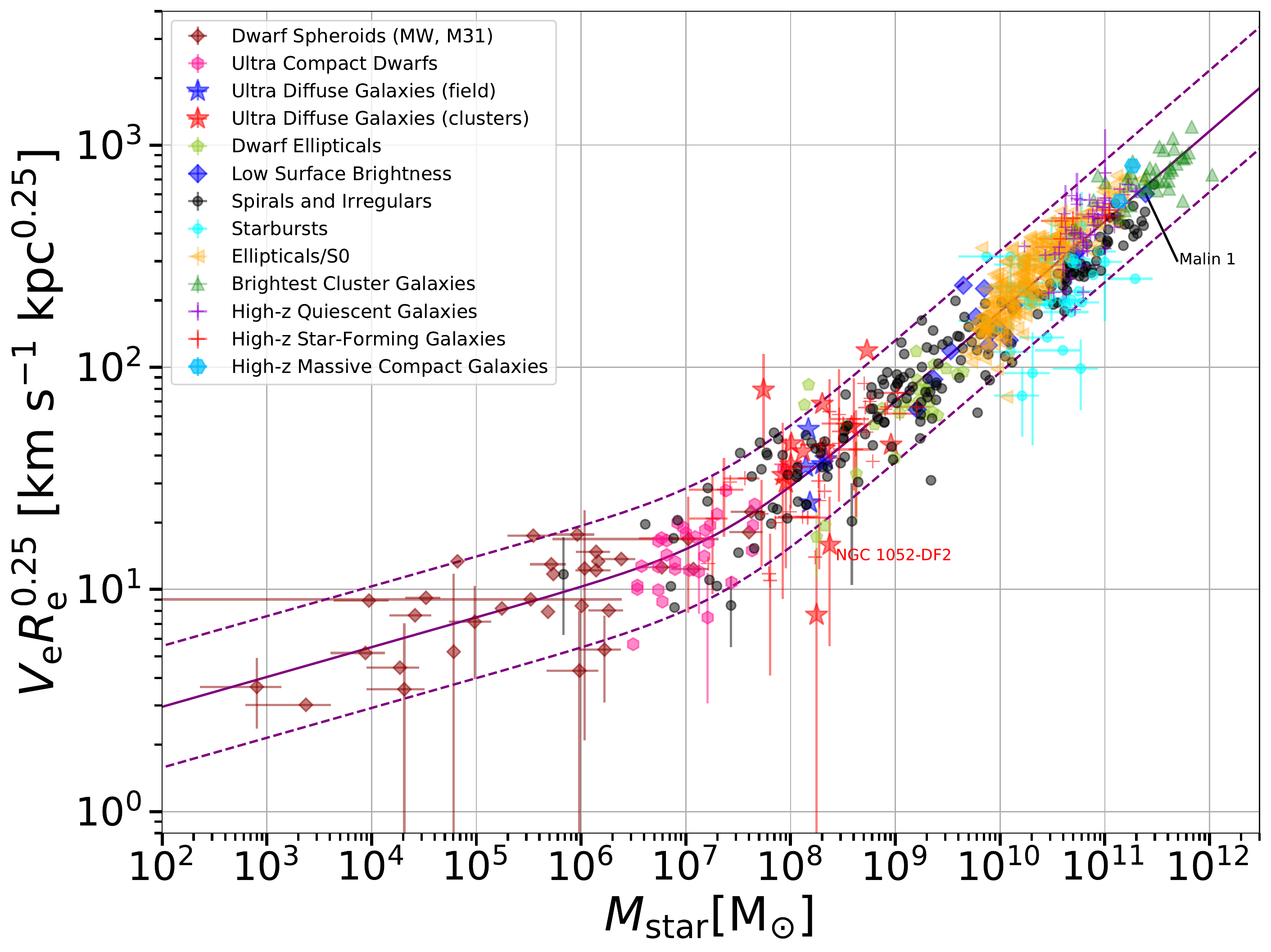}
    \caption{\label{V_R0dot25_Mstar} {\bf A relationship between stellar masses of galaxies ($M_{\rm star}$) and
        the product of the dynamical velocity and size to one fourth power both at the galaxy
        effective radius ($V_{\rm e}R_{\rm  e}^{0.25}$).} Types of
    galaxies are listed in Table~\ref{tab_stats}. $V_{\rm e}R_{\rm  e}^{0.25}$  represents the fourth root of the total binding energy within $R_{\rm  e}$ of a galaxy (see text). The best-fit double power law is shown as a solid curve, while
    two dashed curves indicate $\pm$2$\sigma$. For clarity, error bars that are
    smaller than 0.15 dex are not plotted in the figure.}
\end{center}
\end{figure*}

\begin{figure*}
  \begin{center}
    \includegraphics[scale=0.3]{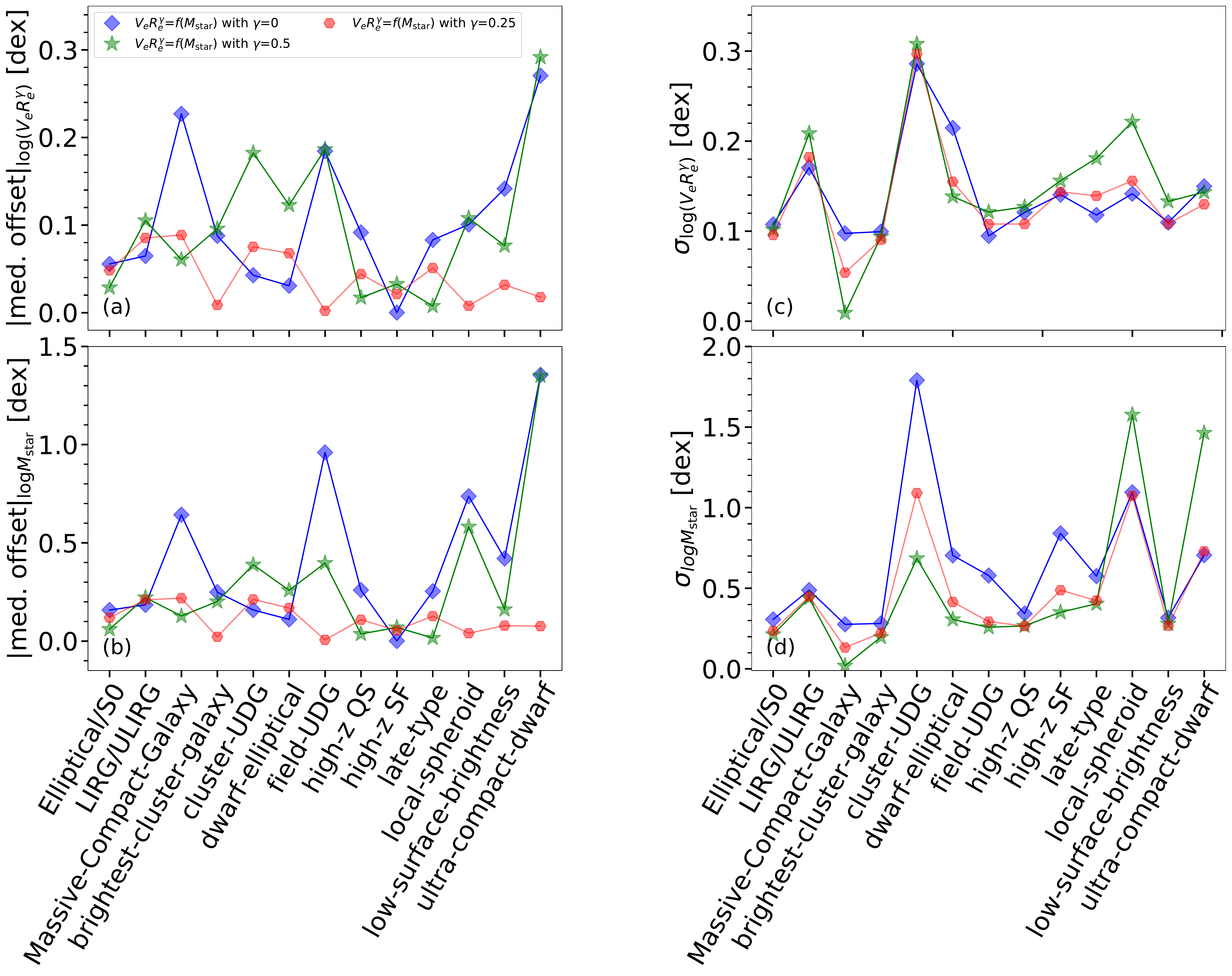}
    \caption{\label{fig_offset_type} {\bf The median offsets and standard deviation of
      individual galaxy types}. {\bf a,} the median offset in
        log($V_{\rm e}R_{\rm e}^{\gamma}$). {\bf b,} the standard deviation in log($V_{\rm e}R_{\rm e}^{\gamma}$). {\bf c,} the median offset in
       log($M_{\rm star}$).  {\bf d,} the standard deviation of in
       log($M_{\rm star}$). }
\end{center}
\end{figure*}

\section{Results}

\subsection{Different combinations of $V_{\rm e}$ and $R_{\rm e}$ in correlations with $M_{\rm star}$}\label{sec_differnt_gamma}

To find the best combination of $V_{\rm e}$ and $R_{\rm e}$ to correlate with $M_{\rm star}$, we employed
the following double-power-law relationship between $V_{\rm       e}R_{\rm        e}^{\gamma}$ and $M_{\rm star}$:
\begin{equation}\label{eqn_dpl_gamma}
V_{\rm       e}R_{\rm        e}^{\gamma}       =       A(M_{\rm star}/M_{0})^{\alpha}(1+M_{\rm   star}/M_{0})^{\beta-\alpha}.
\end{equation}
For each $\gamma$, we carried out the best fitting with the Python code \texttt{scipy.optimize.curve\_fit} and estimated the dispersion of
the derived correlation. Figure~\ref{dispersion_gamma} shows the overall dispersion as a function
of $\gamma$.
Figure~\ref{V_R0dot25_Mstar} gives an example of the above double-power-law relationship for
$\gamma$=0.25.

At different ${\gamma}$, $V_{\rm       e}R_{\rm        e}^{\gamma}$ represents different physical properties of galaxies:
\begin{equation}\label{eqn_gamma_physics}
 V_{\rm       e}R_{\rm        e}^{\gamma}  =
\begin{cases}
   V_{\rm       e}   &  \gamma=0, \\
   V_{\rm       e}R_{\rm        e}^{0.25} \propto (E_{\rm binding,e})^{0.25}     &  \gamma=0.25, \\
   V_{\rm       e}R_{\rm        e}^{0.5} \propto (M_{\rm dyn,e})^{0.5}     &  \gamma=0.5, \\
    \end{cases}     
\end{equation}
where $E_{\rm binding,e}$ is the binding energy within the effective radius and $M_{\rm dyn,e}$ is
the dynamical mass within the effective radius. Here at $\gamma$=0.25 we show that
\begin{equation}
  \begin{split}
    E_{\rm binding,e} &\approx \frac{GM_{\rm dyn,e}^{2}}{R_{\rm e}} \equiv \frac{R_{\rm        e}V_{\rm e}^{4}}{G} \\
    &=4.62\times10^{48} \;{\rm erg} \left( \frac{V_{\rm e}R_{\rm        e}^{0.25}}{\rm km\;s^{-1}\;kpc^{0.25}} \right)^{4}.
  \end{split}
\end{equation}

Figure~\ref{dispersion_gamma}  (a) and (b) show the dispersion
of   log($V_{\rm  e}R_{\rm   e}^{\gamma}$)  and   the  dispersion   of
log($M_{\rm star}$) as a function of $\gamma$, respectively. Two types
of dispersions are measured: one is the standard deviation; another is
weighted  by  the number  of  different  galaxy  types as  $\sigma$  =
(${\sum}\frac{({\rm                                      obs}_{i}-{\rm
    pred}_{i})^{2}}{w_{i}}$/$\sum\frac{1}{w_{i}}$)$^{0.5}$,      where
$w_{i}$ is  the total number  of the galaxy  type that a  given galaxy
belongs  to.  This  latter  dispersion  is to  balance  the fact  that
different galaxy types have different number of objects in this study,
so that  the dispersion  is not  dominated by the  galaxy type  with a
large number  of objects.   In Figure~\ref{dispersion_gamma}  (a), the
minima of $\sigma_{{\rm log}(V_{\rm  e}R_{\rm e}^{\gamma})}$ occur at
$\gamma$=0.21  and 0.22\update\,  for unweighted  and weighted  cases,
respectively.   In Figure~\ref{dispersion_gamma}  (b),  the minima  of
$\sigma_{{\rm log}(M_{\rm  star})}$ are  located at  $\gamma$=0.25 for
both weighted and unweighted cases.

\subsection{A universal correlation between  stellar masses and binding energies of galaxies}

As shown in the above section, the stellar mass of a galaxy is best correlated
with $V_{\rm   e}R_{\rm    e}^{\gamma}$ when $\gamma$ is around 0.25, i.e., the relationship has the minimum dispersion as compared
to other combinations of $V_{\rm   e}$ and $R_{\rm    e}$. Since $V_{\rm   e}R_{\rm    e}^{0.25}$ represents
the fourth root of the binding energy within the effective radius of a galaxy (Equation~\ref{eqn_gamma_physics}), we thus refer this correlation
as the stellar mass-binding energy relationship.
As shown in Figure~\ref{V_R0dot25_Mstar}, all galaxies  together  define
a double  power law function  of
\begin{equation}\label{eqn_binding}
V_{\rm       e}R_{\rm        e}^{0.25} = A(M_{\rm star}/M_{0})^{\alpha}(1+M_{\rm   star}/M_{0})^{\beta-\alpha},
\end{equation}
with    the   best-fitted    $A$=15.7$\pm$2.2\;km\,s$^{-1}$\,kpc$^{0.25}$,
$M_{0}$=(2.5$\pm$1.0)$\times$10$^{7}$\;M$_{\odot}$,    $\alpha$=0.134$\pm$0.020     and
$\beta$=0.406$\pm$0.007\update. The whole relationship has standard deviation of
\scatterLOGVR\, in  log($V_{\rm  e}R_{\rm e}^{0.25}$)  and  \scatterLOGMSTAR\,  in
log($M_{\rm star}$). It covers a stellar  mass range of nine orders of
magnitude, from  one-thousand-solar-mass dwarf spheroidal  galaxies to
the brightest galaxies in clusters with stellar masses of one-thousand-billion solar  masses.  As shown
in  Figure~\ref{fig_offset_type},  the  median offsets  of  individual
galaxy types from the best fit are smaller than the standard deviation
of  the whole  sample.  Individual  galaxy types  have a small  standard
deviation too, except for ultra diffuse galaxies in clusters that have
a standard deviation about two times that of the whole sample. Large observational
errors  account for at least a part of the large dispersion of
these galaxies.

\begin{figure}
  \begin{center}
    \includegraphics[scale=0.35]{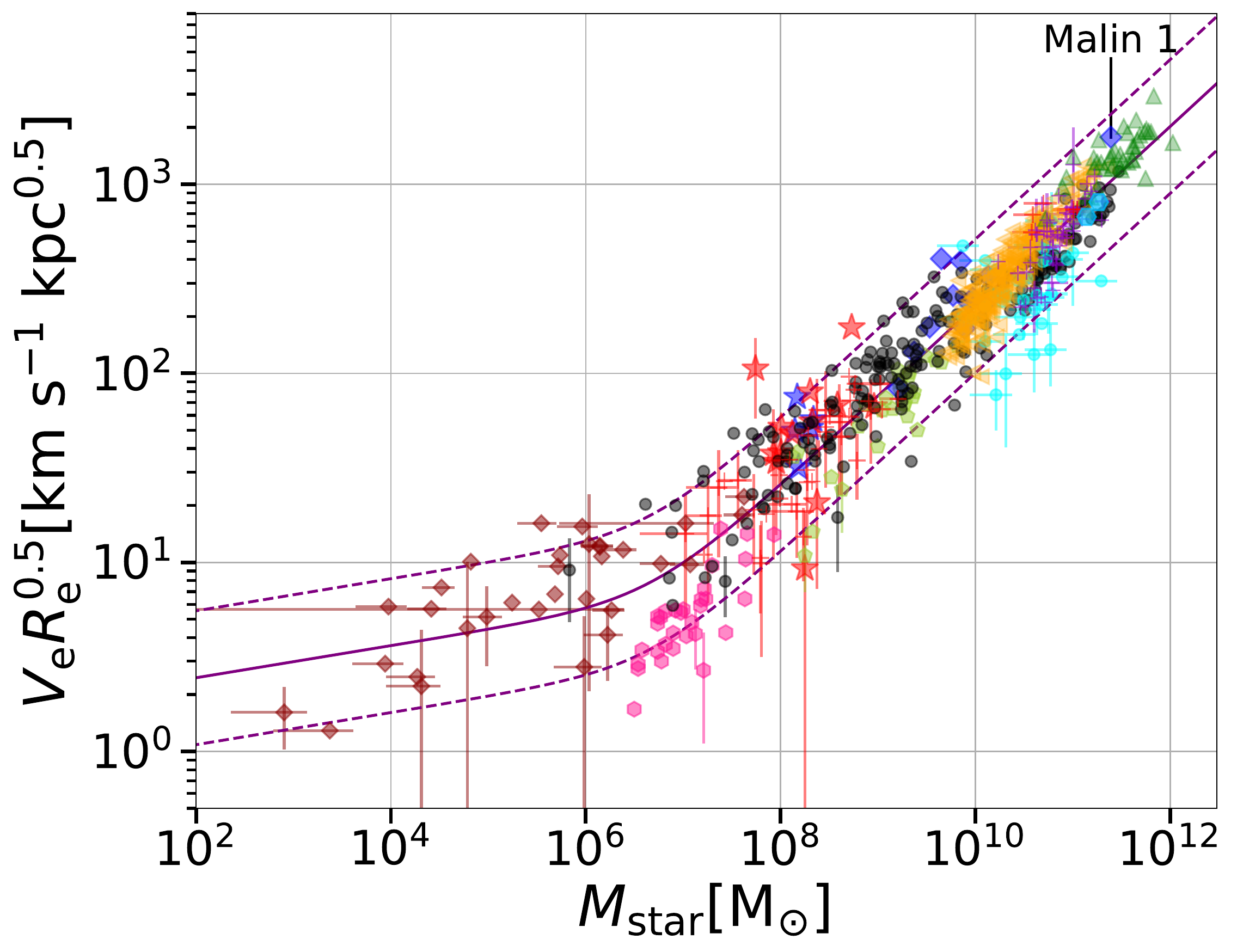}
    \caption{\label{V_R0dot5_Mstar}  {\bf $V_{\rm  e}R_{\rm e}^{0.5}$
       that represents the square root of  the dynamical mass within $R_{\rm e}$,  as a  function  of the galaxy stellar  mass.}
    For clarity, error bars that are
    smaller than 0.15 dex are not plotted in all panels.  Symbols   represent  different   galaxy   types   as  shown   in
      Figure~\ref{V_R0dot25_Mstar}. The best-fit double power law is shown as a solid curve, while   two dashed curves indicate $\pm$2$\sigma$. }    
\end{center}
\end{figure}

Figure~\ref{fig_offset_type}   further  compares  the  stellar
mass-binding  energy  relationship (Equation~\ref{eqn_dpl_gamma}  with
$\gamma$=0.25)  to  the case  with  $\gamma$=0.5  that represents  the
relationship between the  stellar mass and dynamical mass  as shown in
Figure~\ref{V_R0dot5_Mstar}.   The latter relationship has  an overall
standard  deviation  of  0.18  dex\update\,  in  log($V_{\rm  e}R_{\rm
  e}^{\gamma}$)  as shown  in  Figure~\ref{dispersion_gamma}, that  is
30\% larger than the stellar  mass-binding energy relationship.  A few
types of galaxies  show much larger systematic  offsets: ultra compact
dwarfs  show a  median  offset in  ${{\rm log}(V_{\rm  e}R_{\rm
    e}^{\gamma})}$=0.27   dex\update\,  that   is  much   larger  than
-0.018 dex\update\,
in  the   stellar  mass-binding  energy  relationship;   in  addition,
LIRGs/ULIRGs, ultra diffuse galaxies both in
field and clusters, dwarf ellipticals, local spheroids all show median
offsets  in ${\rm  log}(V_{\rm e}R_{\rm  e}^{\gamma})$ larger
than 0.1 dex, while none of a  galaxy type shows such an offset in the
stellar  mass-binding energy  relationship. As  a result,  the stellar
mass-binding energy relationship  should not be caused  by the relationship
between the stellar mass and dynamical mass.

Figure~\ref{fig_offset_type}  also   shows  that  in  the   case  with
$\gamma$=0  several galaxy types   show  median  offsets  in  ${\rm  log}(V_{\rm
    e}R_{\rm e}^{\gamma})$ larger than 0.1 dex, including  massive  compact galaxies,  ultra
diffuse galaxies in  field, low surface brightness  galaxies and ultra
compact  dwarfs. This suggests that
the stellar mass-binding energy relationship is not caused by the
relationship between the stellar mass and the dynamical velocity either.

\begin{figure}
  \begin{center}
    \includegraphics[scale=0.35]{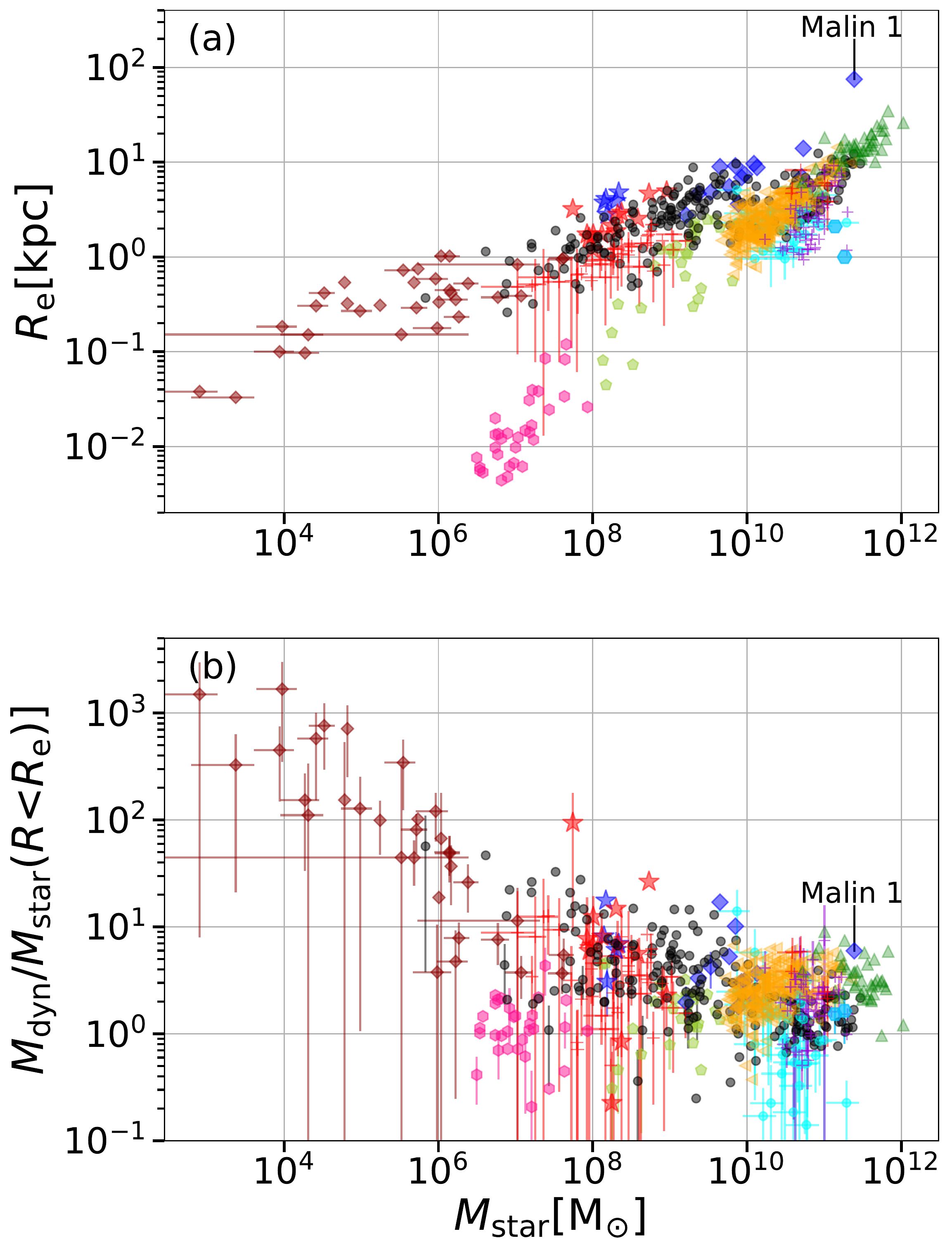}
    \caption{\label{Re_Mstar} {\bf (a) The galaxy effective
        radius as a function of the galaxy stellar mass}. {\bf (b) The dynamical to stellar mass ratio within
        effective radii as a function of the galaxy stellar mass}. For clarity, error bars that are
    smaller than 0.15 dex are not plotted in all panels. Symbols represent different galaxy types as shown in Figure~\ref{V_R0dot25_Mstar}. }
\end{center}
\end{figure}

The scatter  of the  stellar mass-binding energy relation shows  no dependence  on other galaxy
properties.    As  shown  in  Figure~\ref{Re_Mstar} (a),  the
effective  radii  of the  sample  cover  more than four  orders  of
magnitude, ranging  from about 5  pc seen in ultra compact dwarfs to about  70 kpc as seen in one of
the largest galaxies -- Malin 1.  As  shown  in  Figure~\ref{Re_Mstar} (b), the dynamical to stellar mass ratio
within effective radii of the sample covers more than a factor of 10 if excluding local spheroidal galaxies that
show much higher ratios. The environments of the
sample range  from the  dense centers of  galaxy clusters  to isolated
field, all  of which show small  offsets from the best  fit.  The star
formation rate of our sample ranges  from $<$ 0.01 solar mass per year
to more than one hundred solar masses per year as seen in starburst
galaxies \citep{Bellocchi13}.

The  redshift  evolution  of  the  stellar mass-binding energy correlation  is  negligible.
High-redshift objects in our sample cover the redshift range from 0.17
to  about  2.4.   They  are  classified  into  three  types  including
star-forming \citep{Genzel17,Miller14},    quiescent \citep{Belli14}   and
massive  compact  ones \citep{VanDeSande11,vanDokkum09}.   As  shown  in
Figure~\ref{fig_offset_type}  and  listed   in  Table~\ref{tab_stats},  the
median  offsets of  all  three  types are  smaller  than the  standard
deviation of the whole sample.

Our  sample contains  individual objects  that are  outliers of  other
scaling laws of  galaxies.  Ultra diffuse galaxies in  field are found
to systematically deviate  by about 0.35 dex in the  velocity from the
Tully-Fisher  relationship  \citep{ManceraPina19,   Shi21},  but  they
follow  the stellar  mass-binding  energy relationship  with a  median
offset of  only $-0.002$ dex\update.   NGC 1052-DF2, an  ultra diffuse
galaxy  in  cluster  \citep{vanDokkum18}, deviates  from  the  stellar
mass-total halo mass relationship by a factor of 400, while it deviates from
the  stellar mass-binding  energy  relationship by  only  a factor  of
2.5\update.  Star-forming  galaxies at high  redshift \citep{Genzel17}
that lack  dark matter  also obey the  relationship very  well.  Dwarf
ellipticals  define  a  different  slope in  the  Faber-Jackson  plane
\citep{Faber76}  as compared  to the  one defined  by ellipticals  and
ultra compact dwarfs \citep{Drinkwater03, deRijcke05, Chilingarian08},
or  equivalently  showing on  average  about  0.4  dex offset  in  the
velocity  dispersion.  But they  do  follow  the stellar  mass-binding
energy relationship with a median offset of $-0.068$ dex\update.

\subsection{A comparison to the fundamental plane of elliptical galaxies}\label{sec_fp}

\begin{figure}
  \begin{center}
    \includegraphics[scale=0.3]{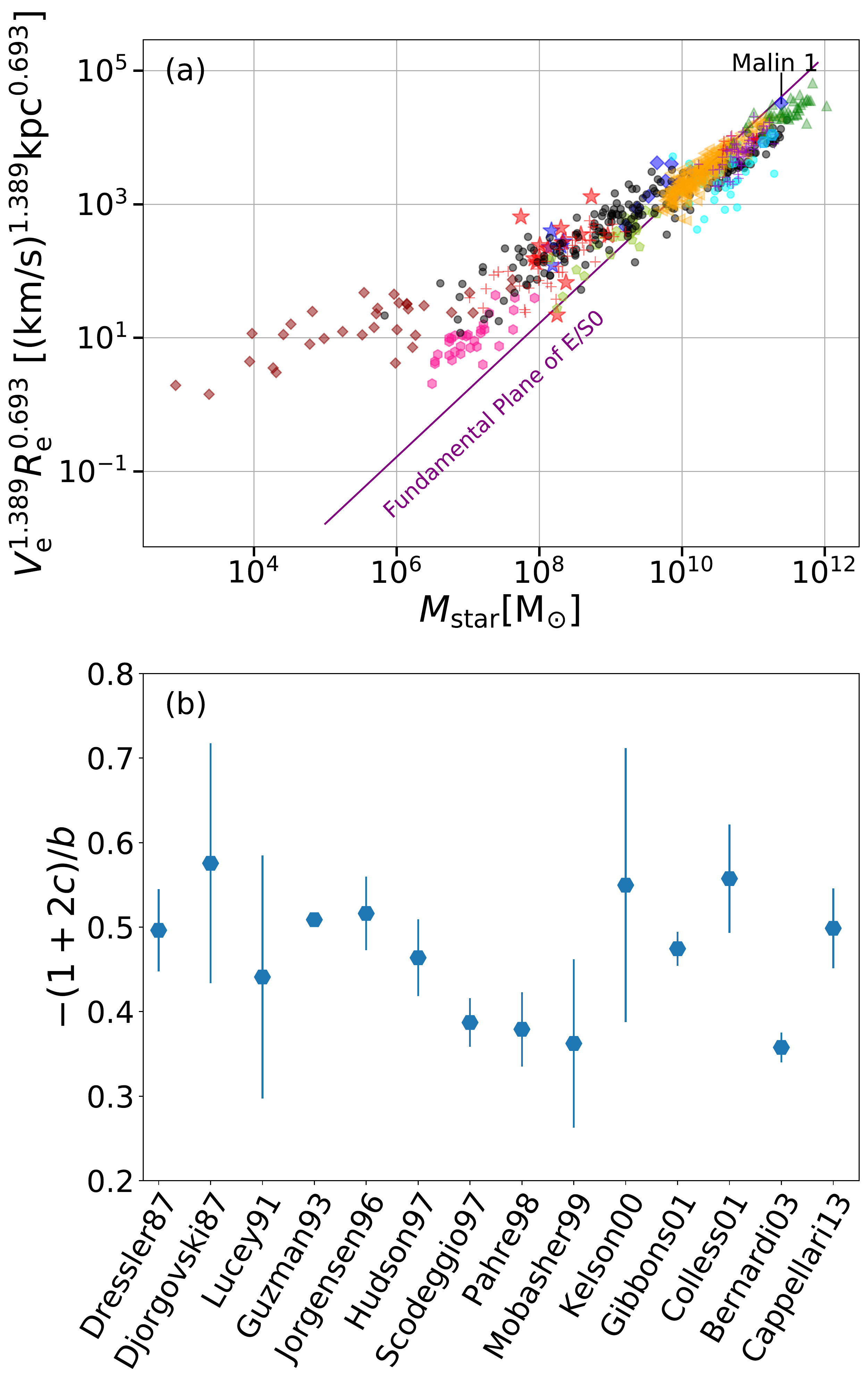}
    \caption{\label{fundamental_plane} {\bf a,} The distribution of all galaxies
      of this study in the fundamental plane of ellipticals. {\bf b,} The distribution of
      $-(1+2c)/b$ in Equation~\ref{eqn_fundamental_plane2} that
      represents $\gamma$  in Equation~\ref{eqn_dpl_gamma} by different studies of ellipticals. }
\end{center}
\end{figure}

\begin{figure}
  \begin{center}
    \includegraphics[scale=0.3]{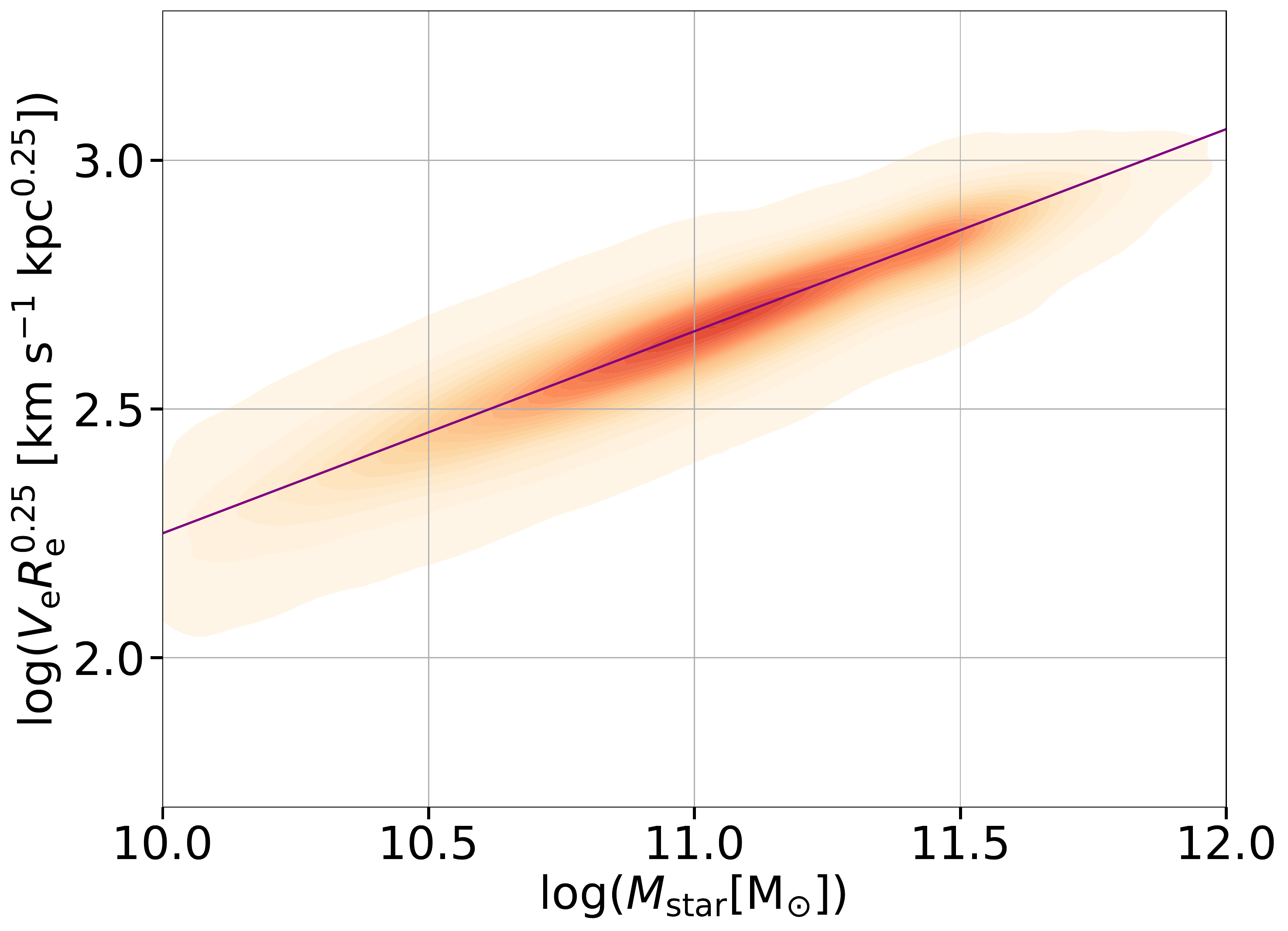}
    \caption{\label{V_R0dot25_Mstar_SDSS} The distribution of $\sim$ 370,000 SDSS early-type galaxies
      in the stellar mass-binding energy relationship. The
      standard  deviation  in
      log($V_{\rm e}R_{\rm e}^{0.25}$) is 0.09 dex\update. The solid line is not the
    fitting to the SDSS data but the best fit from Figure~\ref{V_R0dot25_Mstar}.}
\end{center}
\end{figure}

The fundamental plane of elliptical galaxies invokes the same
three physical parameters as the stellar mass-binding energy relation does
in this study.
The fundamental plane is generally written as
\begin{equation}\label{eqn_fundamental_plane}
{\rm log}R_{\rm 2D,e} = a + b\,{\rm log}\sigma_{0} + c\,{\rm log}\Sigma_{\rm e},
\end{equation}
where $R_{\rm 2D,e}$  is the 2-D effective radius, $\sigma_{0}$ is the central
velocity dispersion and $\Sigma_{\rm e}$ is the effective surface
brightness in optical or near-IR.  The equation can be transformed into
\begin{equation}\label{eqn_fundamental_plane2}
  {M_{\rm star}} \propto V_{\rm e}^{(-b/c)}R_{\rm e}^{(1+2c)/c}= ( V_{\rm e}R_{\rm e}^{-(1+2c)/b} )^{(-b/c)}.
\end{equation}
Here $M_{\rm star}$  represents the stellar light, $V_{\rm  e}$ represents the
$\sigma_{0}$ and 3-D $R_{\rm e}$ represents the $R_{\rm 2D,e}$, all of
which has a  small dispersion that should not affect  the derived power
indices     (see      \S~\ref{sec_samp_gen_proc}).     By     adopting
$b$=1.063$\pm$0.041   and    $c$=$-$0.765$\pm$0.023   as    derived   by
\citet{Cappellari13}   for    ellipticals   used   in    this   study,
Figure~\ref{fundamental_plane} (a)  shows the distribution of  all our
galaxies in  the fundamental plane.  As expected, it  clearly shows
that the  fundamental plane only  holds for ellipticals but  not other
types of galaxies.

It has  been known that  the fundamental  plane of ellipticals  can be
inferred  through   virial  equilibrium,   or  it  is   essentially  a
relationship between the  stellar mass and dynamical  mass. To confirm
this,  Figure~\ref{fundamental_plane} (b)  plots  the distribution  of
$-(1+2c)/b$ that  represents $\gamma$  in Equation~\ref{eqn_dpl_gamma}
from  different studies  of the  fundamental plane  \citep{Dressler87,
  Djorgovski87, Lucey91, Guzman93, Jorgensen96, Hudson97, Scodeggio97,
  Pahre98,  Mobasher99,  Kelson00, Gibbons01,  Colless01,  Bernardi03,
  Cappellari13}. As  shown in the  figure, the $-(1+2c)/b$  is between
0.35 and 0.6,  with a median and standard  deviation of 0.48$\pm$0.07.
That is close to $\gamma{\equiv}$0.5 when Equation~\ref{eqn_dpl_gamma}
represents a relationship between the stellar mass and dynamical mass,
while   significantly   different   from   $\gamma{\equiv}$0.25   when
Equation~\ref{eqn_dpl_gamma}  represents  a relationship  between  the
stellar mass  and binding  energy. We further  fitted our  galaxies by
excluding         local        spheroidal         galaxies        with
Equation~\ref{eqn_fundamental_plane} to  avoid the bending at  the low
mass end through the Python code \texttt{statsmodels.api.OLS}.  It is found that  $a$=2.416$\pm$0.088, $b$=1.569$\pm$0.028,
$c$=-0.645$\pm$0.013 for  the case that $M_{\rm  star}$ represents the
stellar light, $V_{\rm e}$ represents the $\sigma_{0}$ and 3-D $R_{\rm
  e}$ represents the  $R_{\rm 2D,e}$. This gives  $-(1+2c)/b$ of about
0.2, supporting a physical  link between  stellar masses  and binding
energies if including galaxy types in addition to ellipticals.

It is thus  clear that elliptical galaxies show  the smallest dispersion
around $\gamma=0.5$,  while all galaxy  types together define  a tight
sequence around  $\gamma=0.25$. We  further made  a comparison  in the
dispersion   of   log$R_{\rm   e}$  for   ellipticals   when   forcing
$-(1+2c)/b$=0.5           and            $-(1+2c)/b$=0.25           in
Equation~\ref{eqn_fundamental_plane}.   By varying  $c$  over a  large
range  of  -2 to  1,  it  is found  that  the  smallest dispersion  in
log$R_{\rm e}$ of the former, that is 0.10 dex\update, is smaller than
the latter by only 3\%.  We also confirmed a similar difference with a
large  sample of  $\sim$ 370,000  elliptical galaxies.   These objects
were          selected          from         SDSS          value-added
catalog\footnote{https://wwwmpa.mpa-garching.mpg.de/SDSS/DR7/}    with
\texttt{FRACDEV\_R}   $>$   0.8   and   \texttt{MSTAR\_TOT}   $>$   10
\citep{Kauffmann03,  Brinchmann04}. We  used  the  total stellar  mass
\texttt{MSTAR\_TOT}, \texttt{V\_DISP} multiplied  with $\sqrt{5/2}$ as
$V_{\rm e}$  and \texttt{PETROR50\_R}  multiplied with 4/3  as $R_{\rm
  e}$.

As  a summary,  the fundamental  plane of  ellipticals represents  the
relationship between  the stellar  mass and  dynamical mass,  which is
physically different from the stellar mass-binding energy relationship
in this  study. The former  is only  valid for ellipticals,  and its
dispersion in  log$R_{\rm  e}$ is only slightly  smaller 
than the dispersion of ellipticals in the stellar mass-binding energy relationship. 
Figure~\ref{V_R0dot25_Mstar_SDSS}  further shows  the distribution  of
SDSS  early type  galaxies in  the plane  of the  stellar mass vs. the binding
energy. As shown in the figure, they lie well on the best
fit  to   galaxies  in  Figure~\ref{V_R0dot25_Mstar}.    The  standard
deviation  in log($V_{\rm  e}R_{\rm e}^{0.25}$) of Figure~\ref{V_R0dot25_Mstar_SDSS}  is found  to be  0.09
dex\update.

\section{Discussions}

\subsection{Comparisons with other  relationships}
  
Under some  circumstances, the stellar mass-binding energy
relationship  can be  transformed  to a  linear  relationship between  the
galaxy size  and halo size, a  result expected in a  scenario that the
specific  angular  momenta   of  haloes  are  the  same   as  those  of
galaxies \citep{Fall80, Mo98}.  In the  case where the dynamical velocity
is  dominated  by  the  gravity of  dark  matter at the effective radius,  $V_{\rm
  e}$=$V_{200}g(c,x)$=$R_{200}hg(c,x)$, where  $h$ is  Hubble constant
of   100  km   s$^{-1}$  Mpc$^{-1}$, $R_{200}$ is the halo radius in kpc  and  $g(c,x)$=$\sqrt{\frac{{\rm
      ln}(1+cx)-cx/(1+cx)}{x[{\rm ln}(1+c)-c/(1+c)]}}$. For $c$ in the
range  of [3,  30] \citep{Maccio07}  and $x$  in the  range of
[0.001,  0.1]   \citep{Kravtsov13},  it  can  be   shown  that
$g(c,x)$$\approx$1.74$x^{0.42}c^{0.3}$ with standard  deviation of 0.1
dex. From Equation~\ref{eqn_binding} when $M_{\rm  star}$ is much larger than $M_{0}$,
we     got
\begin{equation}
  \begin{split}
  R_{\rm     e}\; & \approx \; 0.014R_{200}[\frac{7.24(M_{\rm star}/
      5{\times}10^{10}M_{\odot})^{0.32}}{R_{\rm
        e}}]^{-2.31}(c/6)^{0.53}  \\
  & \approx\; 0.014R_{200}(c/6)^{0.53},
  \end{split}
\end{equation}
if assuming             $R_{\rm             e}$\,=\,7.24\,kpc\,$(M_{\rm
  star}/5{\times}10^{10}M_{\odot})^{0.31}$ whose  power index  is 20\%
steeper  than the  observation \citep{vanderwel14}.  The  derived galaxy
size-halo size linear  relationship is then quite close to the  one that is
obtained   from  the   abundance  matching \citep{Kravtsov13}.    A  key
difference  is   that  our   estimate  shows   a  dependence   on  the
concentration of the halo.

The Tully-Fisher relationship \citep{Tully77}  of late-type galaxies and
Faber-Jackson  relationship \citep{Faber76} of  early-type galaxies  are
independent of  our stellar  mass-binding  energy correlation as  the
former two are essentially the relationship between the halo mass and stellar
mass. But the latter can be  transformed to the former two by assuming
certain functions of galaxy sizes  with their stellar masses.  For the
Tully-Fisher relationship,  we used the above  conversion from $V_{\rm  e}$  to $V_{200}$  and  found  $M_{\rm star}$  $\propto$
$V_{200}^{3.7}$  \citep{Lelli16},  if   adopting  $R_{\rm  e}$
$\propto$ $M_{\rm star}^{0.35}$ whose power index is 40\% steeper than
the    observation \citep{vanderwel14}.      For    the    Faber-Jackson
law,  the stellar  mass is  found to  be proportional  to the
velocity dispersion at the effective radius  to the power index of 4.0
\citep{Drinkwater03},  if   adopting  $R_{\rm   e}$  $\propto$
$M_{\rm  star}^{0.58}$ whose  power  index is  20\%  smaller than  the
observation \citep{vanderwel14}.

\subsection{A toy model of self regulation}

The tight correlation in  Figure~\ref{V_R0dot25_Mstar} points to the universal
role of  the binding energies of galaxies in the evolution  of stellar
masses. The binding energy within the effective radius may set
a threshold above which stars/gas cannot survive by being expelled out
of  galaxies.    Studies  of  galaxy  formation   and  evolution  have
recognized  many mechanisms  that affect  energies of  gas and  stars.
Some are negative  by increasing the energies such  as ram pressuring,
tidal stripping, supernova feedback etc. Some are positive by removing
the energies such as cooling, turbulence etc.  In a galaxy with a higher
threshold, positive mechanisms work efficiently to accumulate more gas
and stars while  the effects of negative mechanisms  are minimized, so
that a more massive galaxy forms or  survives, and vice versa for a halo
with a lower threshold.

To reproduce  the observed slopes  of the stellar  mass-binding energy
relationship, we present a toy model of self-regulation feedback where
binding  energies of  galaxies  self-balance the  energies of  stellar
systems that have been injected by supernovae feedback over the galaxy
history. We have
\begin{equation}
 E_{\rm binding, e} \approx E_{\rm tot{-}SN}.
\end{equation} 
The supernovae event per time  can be written
as a function of the galaxy stellar mass \citep{Graur15}:
\begin{equation}
 R_{\rm SN} =\frac{dN_{\rm SN}}{dt}\;{\propto}\;M_{\rm star}^{1+B}.
\end{equation}
Supernovae continuously inject energies into
the interstellar media that form stars
over the galaxy history. The deposited energy
is proportional to the number of gas particles that is the gas mass, so that the final energies of stellar systems
injected by supernova are
\begin{equation}\label{eqn_toy_model}
  E_{\rm tot{-}SN} \;{\propto}  \int R_{\rm SN}\,dM_{\rm gas}
  \; {\propto} \int R_{\rm SN}\,({\rm SFR})\,dt
  = \int R_{\rm SN}\,dM_{\rm star}
\end{equation}
Here we assume a linear relationship between the star formation rate (SFR) and gas mass \citep{Gao04} instead of
other relationships \citep{Kennicutt98, Shi14, Shi18}, since only energies that inject into star-forming gas
contribute to the final energies of stellar systems. 
The combination of the above two equations gives
\begin{equation}
E_{\rm tot{-}SN} \; {\propto} M_{\rm star}^{B+2}
\end{equation}
The observations  give $B$ around $-0.5$  for both type I  and type II
supernova    \citep{Graur15},     so    that    we     have   $E_{\rm binding, e}$  ${\propto}$  $M_{\rm  star}^{1.5}$,  which  is
close to the slope (4$\beta$ $\sim$  1.62) we observe above $M_{0}$ as
seen  in   Equation~\ref{eqn_binding}.

As shown in Figure~\ref{V_R0dot25_Mstar},  extremely low mass galaxies
show a different  slope as compared to those above  the $M_{0}$.  This
may be because in these tiny  galaxies the feedback is so efficient so
that   only  a   small   amount   of  gas   and   stars  can   sustain
\citep{Wetzel16}.  A typical $V_{\rm  e}R_{\rm e}^{0.25}$ of these low
mass galaxies  is about 5  km s$^{-1}$ kpc$^{0.25}$, giving  the galaxy binding energy $E_{\rm  binding, e}$ of 2.9$\times$10$^{51}$ erg.
This is a  typical energy of a single supernova  event, meaning that a
single  supernova explosion  may  be powerful  enough  to disrupt  the
galaxy.  In this case,  no integration in Equation~\ref{eqn_toy_model}
gives $E_{\rm tot{-}SN}$  $\propto$ $R_{\rm SN}$, and  so that $E_{\rm
  binding,  e}$ $\propto$  $M_{\rm star}^{0.5}$  whose power  index on
$M_{\rm star}$ is also close to the slope (4$\alpha$ $\sim$ 0.54) that
we observe below $M_{0}$ in Equation~\ref{eqn_binding}.

The  above  self-regulation scenario  requires  a  thorough mixing  in
kinematics among  stars, gas  and dark matters  so that  the initially
released energy into the stellar system  is equal to the final binding
energy of all  particles within the effective radius,  instead of only
the part of stars.

This toy  model emphasizes the  history of the stellar  growth through
which  star-forming gas  continuously  receives  feedback  from
supernovae, while the feedback from active galactic nuclei most likely
plays its major role in terminating  star formation at the final stage  of the stellar
growth. This  may explain  why the  slope of  the relationship  can be
produced without invoking feedback from active galactic nuclei.

It should be pointed out that this toy model illustrates one example about the
interplay between galaxy binding energies and their stellar
masses. The underlying physical mechanisms may be much more
complicated. An in-depth comparison with simulations may give some key
insights.

\subsection{Some potential applications}

The universality of the stellar mass-binding  energy relationship makes it
potential a distance estimator, especially that it is independent of
galaxy types and shows little  redshift evolution.  For stellar masses
above $M_{0}$, the  luminosity distance is given by $D_{\rm  L}$ = 155
Mpc  $(\frac{V_{\rm   e}}{\rm  km/s})^{1.78}(1+z)^{-0.89}M_{\rm  star,
  Mpc}^{-0.72}(\frac{R_{\rm e}}{\rm arcsec})^{0.44}$,  where $M_{\rm star,
  Mpc}$ is  the stellar mass at  a luminosity distance of  1 Mpc.  The
1-$\sigma$   uncertainty  is   0.2   dex   for  individual   distance
estimates. As  compared to  the Tully-Fisher  relationship, it  has an
advantage by measuring the velocity at the effective radius instead of
that at the outer flat part of the rotation curve.

The universality of
the   relationship    also     challenges    the    modified    Newtonian
dynamics \citep{Milgrom83},  which has  a form  of $VR^{0.5}$  $\propto$
$M_{\rm star}^{0.5}$ at the high acceleration regime and $V$ $\propto$
$M_{\rm star}^{0.25}$  at the low  acceleration regime, none  of which
fits Equation~\ref{eqn_binding}.

\section{Conclusions}

In this study, it is found that the galaxy stellar mass is universally
correlated with  $V_{\rm e}R_{\rm  e}^{\gamma}$ when  $\gamma$=0.25.  The
$V_{\rm e}R_{\rm  e}^{0.25}$ represents the  fourth root of  the total
binding energies  of galaxies including  dark matter and  baryons within
the galaxy effective radii.   The stellar mass-binding energy correlation
holds for a variety of galaxy  types and shows a standard deviation of
\scatterLOGVR\,    in    log($V_{\rm    e}R_{\rm    e}^{0.25}$)    and
\scatterLOGMSTAR\,  in  log($M_{\rm   star}$),  with  little  redshift
evolution.  It  is  physically
different from  the fundamental  plane of ellipticals  that represents
the relation between the stellar mass  and dynamical mass. A toy model
of self-regulation between binding energies and supernovae feedback is
proposed to  reproduce the  observed slopes  of the  correlation.  The
relationship can  also be used  to estimate  the distance of  a galaxy
with an uncertainty of 0.2 dex, independent of the galaxy type.

\section*{Acknowledgements}

We thank the referee for  helpful and constructive comments that
improved the paper significantly.
Y.S. acknowledges  the support from  the National Key R\&D  Program of
China (No.  2018YFA0404502, No.  2017YFA0402704), the National Natural
Science  Foundation  of  China  (NSFC grants  11825302,  11733002  and
11773013), and the Tencent Foundation  through the XPLORER PRIZE.
S.M. is partly supported by the National Key Research and Development Program of China (No. 2018YFA0404501), by the National Science Foundation of China (Grants  11821303, 11761131004 and 11761141012). We acknowledge the
science research grants from the China
Manned Space Project with NO. CMS-CSST-2021-B02.

%%%%%%%%%%%%%%%%%%%%%%%%%%%%%%%%%%%%%%%%%%%%%%%%%%
\section*{Data Availability}

All the data used here are available upon reasonable
  request.



\bibliographystyle{mnras}
\bibliography{ms} % if your bibtex file is called example.bib

%\begin{thebibliography}{99}
%\end{thebibliography}

%%%%%%%%%%%%%%%%%%%%%%%%%%%%%%%%%%%%%%%%%%%%%%%%%%

%%%%%%%%%%%%%%%%% APPENDICES %%%%%%%%%%%%%%%%%%%%%

%\appendix



%%%%%%%%%%%%%%%%%%%%%%%%%%%%%%%%%%%%%%%%%%%%%%%%%%

% Don't change these lines
\bsp	% typesetting comment
\label{lastpage}
\end{document}